\documentclass[11pt]{iopart}
\usepackage{graphicx}
\usepackage{psfrag}

\begin{document}
\title{Scaling Cosmologies from Duality Twisted Compactifications}
\date{\today}
\author{James E. Lidsey and Karim A. Malik}
\address{Astronomy Unit, School of Mathematical Sciences\\
  Queen Mary, University of London\\
  Mile End Road, London E1 4NS\\
  United Kingdom}
\eads{\mailto{J.E.Lidsey@qmul.ac.uk, K.Malik@qmul.ac.uk}}
\submitto{JCAP}
\begin{abstract}
Oscillating moduli fields can support a cosmological scaling solution
in the presence of a perfect fluid when the scalar field potential
satisfies appropriate conditions. We examine when such conditions
arise in higher-dimensional, non-linear sigma-models that are reduced
to four dimensions under a generalized Scherk-Schwarz
compactification. We show explicitly that scaling behaviour is
possible when the higher-dimensional action exhibits a global ${\rm
SL}(n,R)$ or ${\rm O}(2,2)$ symmetry. These underlying symmetries can
be exploited to generate non-trivial scaling solutions when the moduli
fields have non-canonical kinetic energy. We also consider the
compactification of eleven-dimensional vacuum Einstein gravity on an
elliptic twisted torus.

\vspace{3mm}
\begin{flushleft}
\end{flushleft}
\end{abstract}
\maketitle

\section{Introduction}

\label{sec:intro}

Scaling solutions play a central role in cosmology 
\cite{w1988,pr1988,crs1997,clw1997,fj1977,ls1998,zws1999,swz1999,nnr2001,clm2004,bcgn2003,hprw2006,ks2006,cnr2005,rrw2007,tw2007,cpr2007,lms1998,mw1998,cmn1998,gl1999}.
They allow one to determine the asymptotic behaviour of various models 
and to establish whether such behaviour is stable or not. They 
also provide a framework for understanding the general dynamics of 
scalar fields in cosmological backgrounds. 
A scaling solution arises when the energy densities 
of the fields that contribute to the total energy density  
vary at the same rate with time. This implies that the 
field densities relative 
to the total density remain constant. 
For a spatially flat and isotropic Friedmann-Robertson-Walker (FRW) 
universe this generates a power-law expansion (or contraction) 
of the scale factor. 

A novel type of scaling behaviour in flat FRW cosmologies
was recently uncovered by 
Karthauser, Saffin and Hindmarsh (KSH) \cite{ksh2007}. 
These authors considered a multi-field scenario consisting of a 
perfect fluid with equation of state $\gamma \equiv (\rho+p)/\rho <2$
and a set of canonical scalar fields interacting through a potential 
\begin{equation}
\label{toypot}
V =\frac{1}{2} e^{c \varphi} \sum_a m_a^2 \chi^2_a  ,
\end{equation}
where $c$ is a constant and $m_a$ represent the masses of the 
moduli fields $\chi_a$. When these fields oscillate 
about the potential minimum, there exists an approximate 
analytical solution where the perfect fluid and scalar fields
maintain constant fractional energy densities given by 
\begin{equation}
\label{fluidscale}
\Omega_{\gamma} = 1- \Omega_s \simeq 1-\frac{3(\gamma -1)}{c^2}  ,
\end{equation}
respectively \cite{ksh2007}. 

Such behaviour may have important implications 
for resolving the moduli problem of the early universe. 
If the oscillating moduli were uncoupled to the $\varphi$-field 
(i.e. if $c=0$),  
they would behave collectively as a pressureless fluid and would 
quickly come to dominate the cosmic dynamics \cite{t1983}. 
On the other hand, massless 
moduli fields behave as a stiff perfect fluid and rapidly 
become subdominant to a fluid such as radiation with $\gamma < 2$. The 
coupling of the $\varphi$-field exponentially suppresses the moduli masses 
in the limit $c\varphi \rightarrow -\infty$ 
in such a way that their energy density remains  
fixed relative to that of the background fluid. 

It is therefore important to consider more general potentials inspired 
by unified field theory that support cosmological scaling 
behaviour driven by oscillating moduli fields. 
The purpose of the present paper is to investigate concrete examples 
that are derived under a generalized dimensional 
reduction scheme introduced by Scherk and Schwarz \cite{ss1979}, 
which is referred to as `reduction with a duality twist' \cite{dh2002}. 
In this scheme, a non-linear sigma-model with global duality symmetry $G$ 
is reduced on a circle (or more generally a torus) 
with periodic coordinate $y \sim y+1$ in such a way 
that the fields are `twisted' over the $S^1$ by an element of 
$G$. The dimensionally reduced theory is 
independent of the compactification coordinate if 
the field dependence on $y$ is of the form 
$\hat{\Phi} (x,y) = \exp (Cy) \phi (x)$, 
where $C$ is a generator of $G$ in an appropriate representation
\cite{c2000,dh2002,hre2005,hre2006,co2006}.  

The structure of the paper is as follows. 
In Section \ref{sec:approx}, we review the derivation of the 
scaling solution and identify the necessary conditions that a 
multi-exponential potential must satisfy for such behaviour to arise. 
Section \ref{sec:twist} develops a compactification 
scheme where a non-linear sigma-model is compactified on a circle 
with a $G$ duality twist and then reduced to four dimensions 
by a standard Kaluza-Klein ansatz. 
We proceed in Sections \ref{sec:noncan} and \ref{sec:U} 
to investigate non-linear sigma-models that exhibit 
global ${\rm SL}(2,R)$ and ${\rm O}(2,2)$ symmetries, respectively, 
and demonstrate that such models can result 
in cosmological scaling. These symmetry groups are motivated by 
the compactification of pure Einstein gravity and the 
Neveu-Schwarz/Neveu-Schwarz (NS-NS) sector of the string effective action
on a two-torus. We then discuss the ${\rm S}(n,R)$ model in Section 
\ref{sec:torus} and  employ our results in Section \ref{sec:ksh}
to resolve an ambiguity of Ref. \cite{ksh2007}. Finally,  
we conclude in Section \ref{sec:discuss}. 
 
Units are chosen such that $16\pi G =1$.

\section{Scaling Cosmology with Oscillating Moduli Fields}

\label{sec:approx}

\subsection{Approximate Analytical Scaling Solution}

The canonical scalar field equations for the system (\ref{toypot}) take the 
form\footnote{The fields are described as canonical when they have standard 
kinetic terms in the action, i.e., when the metric on the 
target (field) space is trivial, $\gamma_{ab} = {\rm diag} (1, \ldots , 1)$. 
The fields are non-canonical when this condition is not satisfied.}
\begin{eqnarray}
\label{toyfield1}
\ddot{\varphi} +3H \dot{\varphi} + \frac{1}{2}c  e^{c\varphi}
\sum_a m_a^2\chi_a^2  =0
\\
\label{toyfield2}
\ddot{\chi}_a+3H\dot{\chi}_a +m^2_a \chi_a e^{c \varphi} =0
\end{eqnarray}
and these fields have a combined energy density 
\begin{equation}
\rho_s= \frac{1}{2} \dot{\varphi}^2 +\frac{1}{2} \sum_a 
\left( \dot{\chi}_a^2  + m_a^2 \chi_a^2  e^{c \varphi} \right) .
\end{equation}

It is assumed that the massive fields $\chi_a$ are oscillating
around the minimum of the potential with a period 
much shorter than the timescales characterized by $H^{-1}$ 
or $\varphi /\dot{\varphi}$. It is further assumed that 
the time average of these oscillations  
decays as $\langle \chi_a^2 \rangle \propto t^{-2\sigma}$ for some constant
$\sigma$ and that the combined energy density in the scalar fields is tracking 
that of the perfect fluid. This implies that the 
universe expands as if it were sourced only by the perfect fluid, 
$H=2/(3\gamma t)$. 

It can then be shown that an approximate solution to Eqs. 
(\ref{toyfield1})-(\ref{toyfield2}) is given by \cite{ksh2007}
\begin{eqnarray}
\label{approxsol}
\varphi \simeq \frac{4 (1- \gamma )}{c \gamma} \ln t  
\\
\chi_a \propto \Re \left[ \frac{1}{t^{\sigma}} \exp \left( -i 
\int^t \omega_a (t') dt' \right) \right] ,
\end{eqnarray}
where $\omega_a \simeq m_a e^{c\varphi /2}$,  
$\sigma = (2-\gamma )/\gamma$ and a time average 
$\langle \dot{\chi}_a^2 \rangle = \omega_a^2 \langle \chi_a^2 \rangle$
has been employed to remove the effects of the oscillations. 
The time-averaged effective energy density of the fields 
therefore scales with cosmic time as 
\begin{equation}
\langle \rho_s \rangle \simeq \frac{8(\gamma -1)}{c^2 \gamma^2} 
\frac{1}{t^2} 
\end{equation}
and this implies that the fraction of the total energy density 
in the scalar fields is given by Eq. (\ref{fluidscale}).  
Hence, the universe exhibits scaling behaviour where  
the energy densities of the fluid and scalar fields 
redshift at the same rate. It is important to emphasize that 
the ratio (\ref{fluidscale}) is independent of the total number  
and masses of the oscillating fields.  
It is also worth remarking that the scaling solution only arises 
for $\gamma >1$ and  $c^2> 3(\gamma -1)$. 

\subsection{Necessary Conditions for Scaling Behaviour}
 
Multi-exponential potentials typically arise in the 
dimensional reduction of supergravity theories. A suitable framework to 
consider, therefore, is given by a set of  
$N$ canonically coupled scalar fields, $\vec{\phi} 
= (\phi_1 , \ldots , \phi_N)$, which interact through 
$M$ exponential terms such that 
\begin{equation}
\label{genpot}
V (\phi) = \sum_{a=1}^M \Lambda_a \exp \left( \vec{\alpha}_a \cdot \vec{\phi}
\right)  ,
\end{equation}
where the constant vectors $\vec{\alpha}_a$ parametrize the field couplings, 
$\vec{\alpha}_a \cdot \vec{\phi} = \sum_{i=1}^N \alpha_{ai}\phi^i$,
and the constants $\Lambda_a$ may be positive or negative. We label 
the $i^{\rm th}$ component of $\vec{\alpha}_a$ by $\alpha_{ai}$. 
The necessary conditions that the potential (\ref{genpot}) must satisfy for 
the existence of the scaling solution (\ref{fluidscale}) 
are that it be semi-positive definite, admit a Minkowski global vacuum and 
contain an overall exponential factor. 

In general, the $\vec{\alpha}_a$ row-vectors form a $M\times N$ matrix, whose 
rank $R$ determines the number of independent $\vec{\alpha}_a$. Models 
of the type (\ref{genpot}) can be separated into two main classes, 
characterized by whether the $\vec{\alpha}_a$-vectors are linearly independent
(Type I) or linearly dependent (Type II) \cite{cnr2005}. 
We focus on the latter class of models in this subsection, 
since the potentials which arise in the following Sections 
belong to this class. Furthermore, if $R<N$, one may perform an 
orthogonal rotation in field space to decouple 
$(N-R)$ of the fields from the potential. 
These degrees of freedom may then be 
consistently set to zero and 
we assume that such a rotation and truncation 
has been performed in what follows. 

A necessary condition for reducing the general type II 
potential (\ref{genpot}) to the form (\ref{toypot}) 
is that there should exist a rotation in field space, 
$\vec{\phi} \rightarrow \vec{\varphi}$, which transforms the potential 
into the separable form 
\begin{equation}
\label{seppot}
V(\varphi ) = e^{c \varphi_1} U(\varphi_2, \ldots , \varphi_N )  ,
\end{equation}
where $c$ is a constant and $U$ is a positive, semi-definite function. 
To establish when this is possible, 
we label the $R$ independent vectors 
$\vec{\alpha}_a$ with $a=1, \ldots , R$ \cite{rrw2007}. 
The remaining dependent vectors $\vec{\alpha}_b$ with 
$R+1 \le b \le M$ can then be expressed as 
linear combinations of the $\vec{\alpha}_a$ vectors such that
$\alpha_{bi} =\sum_{a=1}^R c_{ba}\alpha_{ai}$, where $c_{ba}$
are constant coefficients. Introducing a unit 
vector $\vec{n}$ that satisfies the condition $\vec{\alpha}_a \cdot \vec{n}
=c$ for all $a\le R$ and defining a new basis 
$\vec{\phi} = \varphi_1 \vec{n} +\vec{\varphi}_{\perp}$ then implies 
that $\alpha_{a1} =c$ in this basis for all $a\le R$ 
\cite{cnr2005,hprw2006,rrw2007}. It then follows that  
the 1-components of the $\vec{\alpha}_b$ vectors are given by 
\begin{equation}  
\alpha_{b1} =\sum_{a=1}^R c_{ba}\alpha_{a1} = c \sum_{a=1}^R c_{ba} 
\end{equation}
and consequently, 
the potential (\ref{genpot}) can be factorized into the 
separable form (\ref{seppot}) if the coefficients $c_{ba}$ are 
related through the so-called `affine' condition 
\cite{cnr2005,hprw2006,rrw2007}: 
\begin{equation}
\label{affine}
\sum_{a=1}^R c_{ba} =1  \qquad \forall \qquad  b=R+1, \dots , M  .
\end{equation}

A further  necessary condition for the potential (\ref{seppot}) 
to reduce to Eq. (\ref{toypot}) is that the function $U$ should  
admit a global minimum: 
\begin{equation}
\label{nec2}
U=0 , \qquad  \partial_i U =0 , \qquad \partial_i \partial_j U >0
\end{equation} 
and depend quadratically on the fields $\varphi_a$ for $a\ge 2$ in the 
neighbourhood of this critical point. 
Conditions (\ref{affine}) and (\ref{nec2}) 
arise naturally in duality twisted  
compactifications of higher-dimensional, non-linear 
sigma-models and we focus on such compactifications in the 
following Section. 

\section{Dimensional Reduction with a Duality Twist} 

\label{sec:twist}

In the Kaluza-Klein compactification of 
a $(D+n+1)$-dimensional supergravity theory to 
$(D+1)$ dimensions, the full set of massless
scalar (moduli) fields $\hat{\Phi}^a$ typically 
parametrize a coset space $G/K$, where $K \subset G$ is 
the maximal compact subgroup of a non-compact Lie 
group $G$. (For a review, see, e.g., 
\cite{lwc2000}). The graviton-moduli sector of the  
dimensionally reduced action is given by a
non-linear sigma-model\footnote{The covariant derivative 
on the $(D+1)$-dimensional spacetime is denoted by $\nabla$ and  
the corresponding derivative on the $D$-dimensional spacetime 
is denoted by $\partial$. We distinguish $(D+1)$-dimensional fields 
from $D$-dimensional ones by a hat. The coordinates of the $D$-dimensional 
spacetime are denoted by $x$ unless otherwise stated.}
\begin{equation}
\label{nonlinearsigma}
S=\int d^{D+1}x \sqrt{|\hat{g}|} \left[ \hat{R} 
+ \frac{1}{4} {\rm Tr} \left( 
\nabla \hat{\cal{M}} \nabla \hat{\cal{M}}^{-1} \right) \right]  ,
\end{equation}
where the symmetric moduli matrix $\hat{\cal{M}} (\hat{\Phi}) \in G$ 
determines the metric on the coset space, $ds^2 
= -\frac{1}{4} {\rm Tr} (d \hat{\cal{M}} d \hat{\cal{M}}^{-1})$. 
Action (\ref{nonlinearsigma}) 
is invariant under the global $G$ symmetry transformation
$\hat{g}_{\mu\nu} \rightarrow \hat{g}_{\mu\nu}$ and 
$\hat{\cal{M}} \rightarrow U^T \hat{\cal{M}} U$, where $U\in G$ is a constant 
matrix. 

The ansatz for a consistent dimensional reduction of action 
(\ref{nonlinearsigma}) on a circle with a $G$ duality twist 
is \cite{ss1979,mo1998,c2000,dh2002,co2006} 
\begin{equation}
\label{higherdimmetric}
d\hat{s}^2_{D+1} = e^{-2\alpha \psi} ds^2_D 
+ e^{2 (D-2) \alpha \psi} dy^2 
\end{equation}
\begin{equation}
\label{Mansatz}
\hat{\cal{M}} (x,y) = \lambda^T (y) {\cal{M}}(x) \lambda (y) 
, \qquad \lambda (y) \equiv  \exp \left( Cy \right)  ,
\end{equation}
where the field $\psi$ parametrizes the radius of the circle and 
\begin{equation}
\label{defalpha}
\alpha \equiv \frac{1}{\sqrt{2(D-1)(D-2)}}   .
\end{equation}
The matrix $C = \lambda^{-1} \nabla_y \lambda =   
- (\nabla_y  \lambda^{-1} )  \lambda$ 
is an element of the Lie algebra of $G$  and has dimensions of mass. 
The map $\lambda (y) = \exp (Cy)$ 
has a monodromy ${\cal{C}} = \exp (C) \in G$ and 
the physically distinct reductions are characterized 
by the conjugacy classes of this monodromy \cite{h1998}. 
The scalar kinetic term in Eq. (\ref{nonlinearsigma}) 
generates both a kinetic term and a scalar field potential in 
$D$ dimensions and the resulting action is given by 
\cite{ss1979,mo1998,c2000,dh2002,co2006}  
\begin{eqnarray}
\label{twistedaction}
S=\int d^Dx \sqrt{|g|} \left[ R -\frac{1}{2} \left( \partial 
\psi  \right)^2 + \frac{1}{4} {\rm Tr}
\left( \partial {\cal{M}} \partial {\cal{M}}^{-1} \right) 
-V (\psi, {\cal{M}} ) \right]
\\
\label{twistedpot}
V \equiv  e^{-2(D-1)\alpha \psi} U ({\cal{M}} ) , \qquad 
U ( {\cal{M}}) \equiv  \frac{1}{2} {\rm Tr} \left( 
C^2 +C{\cal{M}}^{-1}C^T{\cal{M}} \right)  .
\end{eqnarray}  

When $D>4$, we may perform a standard Kaluza-Klein
dimensional reduction on an isotropic 
$(D-4)$-dimensional torus $T^{D-4}$ with periodic coordinates 
$z^a \sim z^a +1$, where 
all the moduli fields on the torus are frozen 
with the exception of the breathing mode, $\sigma$, which parametrizes 
the volume of $T^{D-4}$. In this case, the metric ansatz is 
\begin{eqnarray}
ds^2_D = e^{2 \epsilon \sigma} ds_4^2 +e^{2 \omega \sigma} \delta_{ab}
dz^adz^b
\nonumber \\
\epsilon \equiv  -\frac{1}{2} \sqrt{\frac{D-4}{D-2}} , \qquad  
\omega \equiv  - \frac{2\epsilon}{D-4}
\label{kkreduce}
\end{eqnarray}
and this results in the four-dimensional action  
\begin{eqnarray}
\label{4twistedaction}
S=\int d^4x \sqrt{|g|} \left[ R - 
\frac{1}{2} \left( \partial \psi \right)^2 -\frac{1}{2}
\left( \partial \sigma \right)^2 + \frac{1}{4} {\rm Tr}
\left( \partial {\cal{M}} \partial {\cal{M}}^{-1} \right)
\right. 
\nonumber \\
\left. 
- e^{-2(D-1)\alpha \psi + 2\epsilon
\sigma} U ({\cal{M}}) \right]  .
\end{eqnarray}
An ${\rm SO}(2)$ rotation on the 
moduli fields $\{ \psi , \sigma  \}$ defined by    
\begin{eqnarray}
\varphi = - \frac{1}{\sqrt{3}} \left[ \sqrt{\frac{2(D-1)}{D-2}} \psi 
+ \sqrt{\frac{D-4}{D-2}} \sigma \right] 
\nonumber 
\\
\gamma = \frac{1}{\sqrt{3}} \left[ \sqrt{\frac{D-4}{D-2}} \psi 
- \sqrt{\frac{2(D-1)}{D-2}}  \sigma \right]  
\label{defy}
\end{eqnarray}
then decouples the $\gamma$-field from the moduli 
potential (\ref{4twistedaction}). Consequently, the potential 
reduces to the separable form required for scaling cosmology: 
\begin{equation}
\label{compactpot}
V= \frac{e^{\sqrt{3} \varphi}}{2}  
{\rm Tr} \left( C^2 +C{\cal{M}}^{-1}C^T{\cal{M}} \right)  .
\end{equation}

The question 
that now arises is whether the function $U({\cal{M}})$ has a 
stable minimum at $U=0$. To proceed, it proves convenient \cite{dh2002} to 
introduce the real vielbein matrix 
${\cal{V}} \in G$ defined by ${\cal{M}} \equiv  
{\cal{V}}^T{\cal{V}}$ and the real, symmetric matrix $Y \equiv 
[\tilde{C} + \tilde{C}^T ]$, where $\tilde{C} \equiv {\cal{V}} C 
{\cal{V}}^{-1}$. The potential (\ref{compactpot}) can then 
be expressed as  
\begin{equation}
\label{morecompactpot}
V = \frac{e^{\sqrt{3} \varphi}}{4} {\rm Tr} ( Y^2 )
\end{equation}
and, since $Y$ is diagonizable and has real eigenvalues, it follows that 
${\rm Tr}(Y^2)$ is the sum of the squares of these eigenvalues. 
Hence, the potential is semi-positive definite, $V \ge 0$, and 
can vanish at a point $\Phi_0$ in the 
moduli space if and only if $Y=0$ \cite{dh2002}.  
This occurs when $\tilde{C}$ is given by 
a rotation generator at that point, $\tilde{C}_0 = - \tilde{C}_0^T$,
which implies that 
\begin{equation}
\label{rotgen}
{\cal{M}}_0 C =-C^T {\cal{M}}_0  .
\end{equation}
The solution to Eq. (\ref{rotgen}) is given by $C= S_0^{-1} R S_0$ and 
${\cal{M}}_0 = S_0^TS_0$, where $S_0 \in G$ is a constant matrix and 
$R=-R^T$ is a generator of  ${\rm O}(n)$. In particular, 
when $C$ is itself a $n \times n$ antisymmetric matrix 
Eq. (\ref{rotgen}) is solved by 
${\cal{M}}_0 = {\rm I}_n$, where ${\rm I}_n$ is the identity 
matrix. 

To summarize thus far, the potential (\ref{compactpot}) 
has a stable Minkowski minimum when 
Eq. (\ref{rotgen}) is satisfied. 
Moreover, the exponential coupling $c=\sqrt{3}$ 
of the $\varphi$-field  to the other moduli is independent of 
both the global symmetry group $G$ and the spacetime dimensionality 
$D$. However, in general the fields parametrized by the 
moduli matrix ${\cal{M}}$ will not necessarily have a canonical kinetic 
energy term, as was assumed in the derivation 
of Eq. (\ref{fluidscale}). Thus, we can not yet 
conclude that scaling cosmologies in four dimensions will necessarily arise. 
On the other hand, if the canonical condition is satisfied, 
we arrive at the generic prediction 
that the scaling cosmologies will be characterized by the 
fractional energy densities 
\begin{equation}
\label{genericpredict}
\Omega_{\gamma} = 2-\gamma , \qquad 
\Omega_s = \gamma -1
\end{equation}
for the fluid and scalar fields, respectively. 

Condition (\ref{rotgen}) implies that 
${\rm Tr} \, C =0$, in which case $C$ belongs to the Lie
algebra of ${\rm SL}(n,R)$. In the following Section, therefore, 
we investigate when scaling cosmologies 
may arise after compactification with a $G={\rm SL}(2,R)$ duality twist. 

\section{Scaling Cosmologies from a ${\mathbf{G=SL(2,R)}}$ Duality Twist}

\label{sec:noncan}

The Kaluza-Klein reduction of $(D+3)$-dimensional 
pure gravity with an Einstein-Hilbert action 
on a two-torus $T^2$ with real periodic 
coordinates $z^a \sim z^a +1$ and fixed volume results 
after truncation to the zero-mode sector in a $(D+1)$-dimensional 
action of the form (\ref{nonlinearsigma}), 
where the scalar fields take values in the coset space
${\rm SL}(2,R)/{\rm SO}(2)$. 
This internal symmetry of the torus can be promoted to 
an external symmetry of the $(D+1)$-dimensional theory, 
which can then be exploited to 
perform a further reduction onto a circle with an 
${\rm SL}(2,R)$ duality twist. 

The ${\rm SL}(2,R)/{\rm SO}(2)$ 
non-linear sigma-model is parametrized by 
a complex `dilaton-axion' field $\hat{\tau} \equiv \hat{\tau}_1 + 
i \hat{\tau}_2= \hat{\chi} +ie^{-\hat{\phi}}$, 
where the moduli matrix is given by 
\begin{equation}
\hat{\cal{M}} = \frac{1}{\hat{\tau}_2} \left( 
\begin{array}{cc}
1 & \hat{\tau}_1 \\
\hat{\tau}_1       &  | \hat{\tau}|^2 
\end{array}
\right)  .
\end{equation} 
The corresponding action (\ref{nonlinearsigma}) has the form    
\begin{eqnarray}
\label{IIBaction}
S =\int d^{D+1} x \sqrt{|\hat{g}|}  \left[ \hat{R} - \frac{1}{2} 
\left( \nabla \hat{\phi} \right)^2 - \frac{1}{2}e^{2\hat{\phi}} 
\left( \nabla \hat{\chi} \right)^2 \right]  .
\end{eqnarray}

In general, a dimensional reduction with a duality twist 
is characterized by the conjugacy classes of the monodromy
\cite{h1998}. The group 
${\rm SL}(2,R)$ has three such classes and it can be shown 
that the lower-dimensional potential admits a stable Minkowski minimum 
only for the elliptic class represented by the monodromy matrix \cite{dh2002}
\begin{eqnarray}
\label{Celliptic}
{\cal{C}} = \left( \begin{array}{cc}
\cos m & \sin m \\
-\sin m       &  \cos  m
\end{array} \right) 
\nonumber \\ 
C= -m J , \qquad 
J \equiv \left( \begin{array}{cc}
0 & 1  \\
-1 &  0 
\end{array}
\right)  ,
\end{eqnarray}
where $m$ is a constant mass parameter and 
$J$ is the generator of ${\rm SO}(2)$. In this case, 
the moduli potential (\ref{twistedpot}) simplifies to \cite{c2000,h2002}
\begin{equation}
\label{fullpot}
U = \frac{m^2}{2} {\rm Tr} \left( -{\rm I}_2 + {\cal{M}}^2 \right) ,
\end{equation}
where the condition ${\cal{M}}^{-1} = -J {\cal{M}} J$ 
has been employed. 

The reduction of theory (\ref{IIBaction})
on $S^1$ with an elliptic ${\rm SL}(2,R)$ duality twist, 
followed by the Kaluza-Klein 
toroidal compactification outlined in Eqs. (\ref{kkreduce})-(\ref{defy}),
therefore leads to the four-dimensional action 
\begin{eqnarray}
\label{fullaction}
S= \int d^4 x \sqrt{|g|} \left[ R - \frac{1}{2} (\partial \varphi )^2 
- \frac{1}{2} (\partial \phi )^2
-\frac{1}{2} e^{2\phi} (\partial \chi )^2   \right.
\nonumber \\
\left. - \frac{m^2}{2}
e^{\sqrt{3} \varphi} \left[
e^{2\phi} \left( \chi^2 +1 \right)^2 + e^{-2\phi} +2 \left( 
\chi^2 -1 \right) \right]  \right]  .
\end{eqnarray}
The potential in Eq. (\ref{fullaction}) 
has a stable Minkowski minimum at $\phi = \chi =0$ and, since the 
axion field $\chi$ appears quadratically in the potential, 
it may consistently be set to zero. In this case, the potential simplifies 
to\footnote{The potential (\ref{potnoaxion}) 
can also be derived by compactifying 
vacuum Einstein gravity on the three-dimensional Bianchi-type 
${\rm VII}_0$ manifold corresponding to the group ${\rm ISO}(2)$   
\cite{bcgn2003,cnr2005,co2006}. 
This follows because the elements of the mass matrix $C$ in 
Eq. (\ref{Celliptic}) correspond to the structure constants of the 
${\rm iso} (2)$ algebra.}  
\begin{equation}
\label{potnoaxion}
V = 2m^2  e^{\sqrt{3} \varphi} {\rm sinh}^2 \phi  
\end{equation}
and Eq. (\ref{potnoaxion}) reduces to Eq. (\ref{toypot}) 
in the limit where $| \phi | \ll 1$. Moreover, 
since the $\phi$- and $\varphi$-fields 
have canonical kinetic energy, we may conclude that they   
drive a scaling cosmology in the presence of a background 
fluid. 

\begin{figure}[!t]
\psfrag{varphi}[][l]{$\varphi$}
\includegraphics[width=14cm]{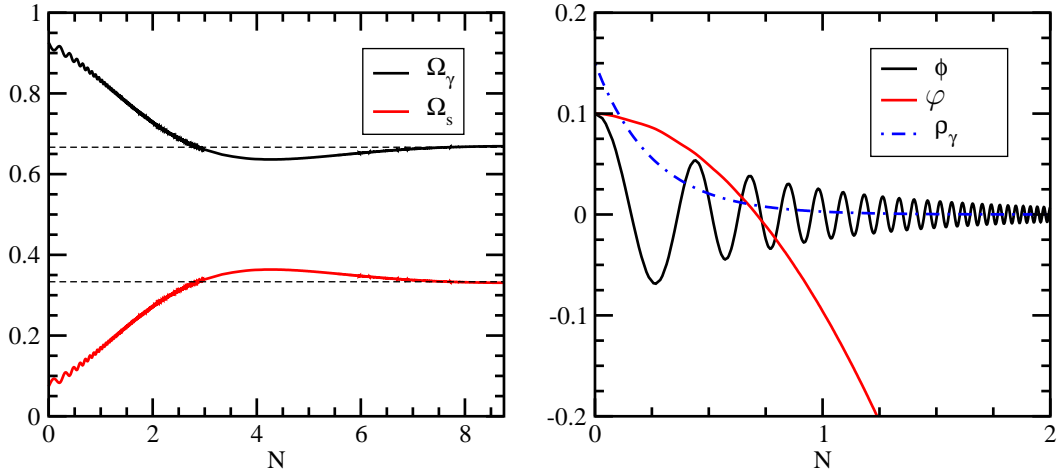}

\caption[] {Illustrating the 
scaling solution for the four-dimensional cosmology  
derived from compactification of the ${\rm SL}(2,R)$ non-linear 
sigma-model with a duality twist in the elliptic conjugacy class. 
The axion field in action (\ref{fullaction}) has been set to zero and 
the background fluid has a relativistic equation 
of state, $\gamma =4/3$. In the left-hand figure, the fractional energy 
densities of the fluid, $\Omega_{\gamma}$, and 
the scalar fields, $\Omega_s$, 
asymptote to the values predicted by the analytical solution (\ref{approxsol}).   
Initial conditions are chosen so that $\rho_{\gamma ,i}=0.15$, 
$\varphi_i = 0.1$, $\phi_i = 0.1$ and 
$\dot{\rho}_{\gamma, i} = \dot{\varphi}=\dot{\phi} =0$ and the mass parameter 
$m=1$. 
The right-hand figure shows the evolution of  the fields with 
respect to the number of e-foldings, $N = \ln  a$. 
}
\label{combined1}
\end{figure}

Fig. \ref{combined1} gives the results of a numerical 
calculation for the system (\ref{potnoaxion}) when  
a relativistic fluid is present. The full field equations are presented 
in Appendix A and follow by specifying $\chi =0$ in Eqs. 
(\ref{appendix1})-(\ref{appendix5}). 
The evolution of the fractional energy densities
for the fluid and scalar fields is 
shown as a function of the number of e-foldings, $N=\ln a$, in the left-hand 
figure. The dashed lines correspond to the values $\Omega_{\gamma} = 2/3$ and 
$\Omega_s =1/3$ that are predicted by the approximate analytic solution 
(\ref{approxsol}) and it is seen that the fractional densities 
asymptote to these constant values after a few e-foldings have elapsed. 
The right-hand figure illustrates the evolution of the respective 
fields. 

The question that now arises is whether similar 
scaling behaviour is possible when the axion field $\chi$ is non-trivial. 
The non-canonical coupling of this field to $\phi$ may 
violate one or more of the necessary conditions that must be 
satisfied for cosmological scaling to proceed.   
We address this question by employing symmetry arguments. Although 
the kinetic sector of these fields is invariant under a 
general ${\rm SL}(2,R)$ transformation, the 
potential in the action (\ref{fullaction}) breaks this symmetry. 
On the other hand, the potential (\ref{fullpot}) is invariant under 
a global ${\rm SO}(2)$ symmetry. This implies that 
four-dimensional action (\ref{fullaction}) exhibits 
an ${\rm SO}(2)$ symmetry: 
\begin{equation}
\label{SO2transform}
\tilde{g}_{\mu\nu} = g_{\mu\nu} , \qquad 
\tilde{\varphi} = \varphi  , \qquad
\tilde{{\cal{M}}} =  \Sigma^T {\cal{M}} \Sigma ,
\end{equation}
where 
\begin{equation}
\label{defSigma}
\Sigma \equiv \left( 
\begin{array}{cc}
\cos \theta  & \sin \theta \\
-\sin \theta &  \cos \theta 
\end{array}
\right)
\end{equation}
is an arbitrary, constant ${\rm SO}(2)$ matrix  
satisfying $\Sigma^T \Sigma = {\rm I}_2$. The transformation 
(\ref{SO2transform}) acts non-linearly on the scalar fields $\phi$ and 
$\chi$ such that 
\begin{eqnarray}
\label{newsol}
e^{\tilde{\phi}} = s^2 e^{-\phi} +(c+s\chi )^2 e^{\phi} 
\nonumber \\
\tilde{\chi} e^{\tilde{\phi}} 
= cs e^{-\phi} - (s-c\chi)(c +s\chi)e^{\phi}  ,
\end{eqnarray}
where $c \equiv \cos \theta$ and $s \equiv \sin \theta$.

We may employ Eq. (\ref{newsol}) 
to generate a cosmological solution involving 
a non-trivial axion field $(\tilde{\chi} \ne 0)$
directly from a background where such a field 
is trivial $(\chi =0 )$. Moreover, the spacetime metric 
is invariant under the action of Eq. (\ref{SO2transform}). 
This implies that a perfect fluid source may be introduced 
into the system without breaking the symmetry of the field equations. Since 
the evolution of this fluid depends only on the cosmic scale factor, 
its energy density and pressure will remain invariant under a global
${\rm SO}(2)$ transformation. Hence, any scaling behaviour 
which arises in the absence of the axion field 
will also be possible when this field is dynamically non-trivial. 
In general, the evolution of the moduli fields $\tilde{\phi}$ and 
$\tilde{\chi}$ will be determined by 
Eq. (\ref{newsol}), where $\chi =0$ and the time-dependence of $\phi$ is 
given by the analytic solution (\ref{approxsol}). 
Although the fractional energy densities of the individual fields will 
alter under the symmetry transformation (\ref{newsol}), 
their combined fractional density 
will remain invariant and will be given by 
$\Omega_s = \gamma -1$. 

\begin{figure}[!t]
\psfrag{varphi}[][l]{$\varphi$}
\includegraphics[width=12cm]{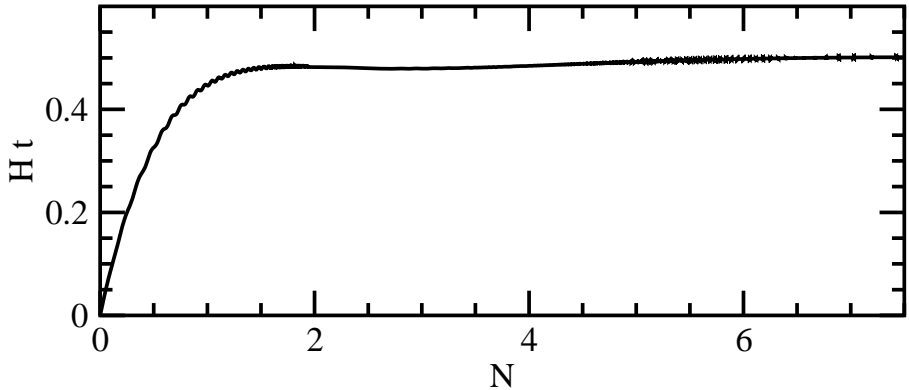}

\caption[]{Illustrating the scaling behaviour  
of the four-dimensional cosmology  
derived from the compactification of the ${\rm SL}(2,R)$ non-linear 
sigma-model with a duality twist in the elliptic conjugacy class.  
The axion field in action (\ref{fullaction}) is non-trivial.  
The background fluid has a relativistic equation 
of state, $\gamma =4/3$, and the initial conditions 
are the same as in Fig. \ref{combined1} with $\chi = 0.1$ 
and $\dot{\chi} =0$. The product $Ht$, where $H$ denotes the 
Hubble parameter, tends to a constant, $Ht \rightarrow 1/2$,  
which implies that the scale factor grows as 
$a \propto t^{1/2}$.  
}
\label{powerlawfigure}
\end{figure}

\begin{figure}[!t]
\psfrag{varphi}[][l]{$\varphi$}
\includegraphics[width=14cm]{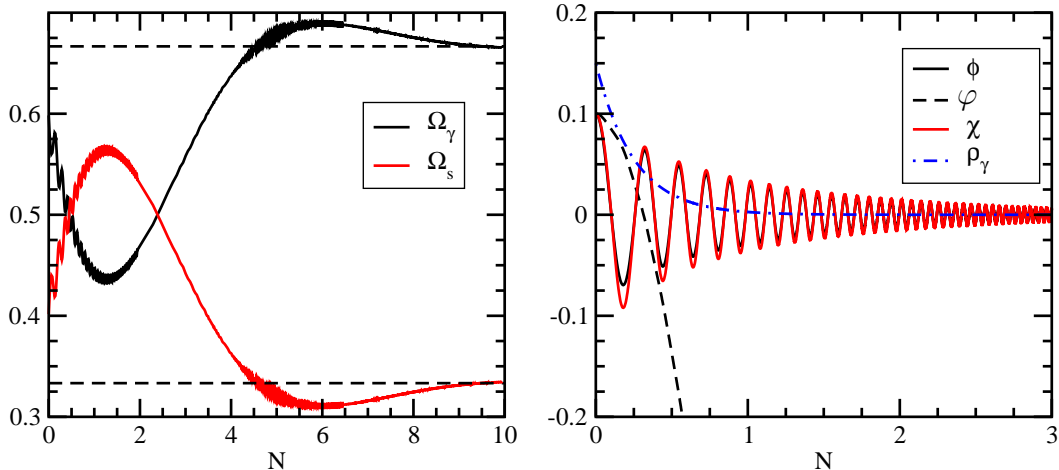}

\caption[] {Illustrating the 
evolution of the fractional energy densities of the fluid and moduli 
fields for the model shown in Fig. \ref{powerlawfigure} 
where the axion field is non-trivial. 
The densities asymptote to the analytically predicted values
$\Omega_{\gamma} = 2/3$ and $\Omega_s =1/3$, respectively.   
The right-hand figure shows the evolution of the fields.  
}
\label{combined2}
\end{figure}

Fig. \ref{powerlawfigure} shows a numerical 
integration of the spatially flat FRW field equations 
(\ref{appendix1})-(\ref{appendix5}) 
derived from the full action (\ref{fullaction}) when a relativistic fluid 
is present. The solution asymptotes to a power-law 
expansion of the scale factor, $a \propto t^{1/2}$. 
Fig. \ref{combined2} shows the evolution of the fluid and moduli field
energy densities and it is seen that these parameters also asymptote to 
their analytically predicted values. 

Thus far, we have considered the cosmological dynamics 
of a model derived from pure Einstein gravity in $(D+3)$ dimensions
compactified on a 2-torus. In the following Section, we extend this 
analysis to consider the non-linear sigma-model that is motivated 
by the compactification of the NS-NS sector of the string effective 
action on $T^2$. 

\section{Scaling Cosmologies from a G=O(2,2) Duality Twist}

\label{sec:U}

The NS-NS sector of the ten-dimensional string effective action 
consists of the graviton, the dilaton and an antisymmetric 
two-form potential, $B$. After toroidal compactification on $T^2$, the 
resulting massless scalar fields parametrize the 
${\rm O}(2,2)/{\rm O}(2) \times {\rm O}(2)$ coset when the value of the 
dilaton field is fixed \cite{ms1993}. In general, 
the group ${\rm O}(n,n)$ is the non-compact, 
pseudo-orthogonal group in $2n$ dimensions (see, e.g., 
\cite{gpr1994}). Its representation is given by matrices $U$ 
that preserve the bilinear form $\eta$: 
\begin{equation}
\label{onn}
U^{\rm T} \eta U =\eta , \qquad 
\eta \equiv \left( \begin{array}{cc} 
0 & {\rm I}_n \\
{\rm I}_n & 0 \end{array} \right)  ,
\end{equation}
where ${\rm I}_n$ is the $n \times n$ identity matrix. 
Since $\eta^2 ={\rm I}_{2n}$, the inverse of $U$ is given 
linearly by $U^{-1} =\eta U^{\rm T} \eta$. 

The group $G= {\rm O}(2,2)$ is 
isomorphic to ${\rm SL}(2,R)\times {\rm SL}(2,R)$ and is 
related to the $T$-duality group of the two-torus \cite{gpr1994}.
This implies that the four degrees of freedom may be arranged into two complex
coordinates defined by \cite{blsw1993}
\begin{eqnarray}
\label{taurho}
\tau \equiv \tau_1 +i\tau_2 = \frac{\Gamma_{mn}}{\Gamma_{nn}}
+i \frac{\sqrt{\Gamma}}{\Gamma_{nn}} \nonumber \\
\rho \equiv \rho_1 +i\rho_2 = B_{mn} +i 
\sqrt{\Gamma}  ,
\end{eqnarray}
where the metric on $T^2$ is denoted by $ds_2^2 = 
\Gamma_{ab} dz^a dz^b$ and has  
determinant $\Gamma \equiv {\rm det \Gamma_{ab}}$.  
The corresponding non-linear sigma-model is given by 
Eq. (\ref{nonlinearsigma}), where the moduli matrix takes the form 
\cite{gpr1994}
\begin{equation}
\label{matrixM1}
{\cal{M}}=\frac{1}{\tau_2\rho_2} 
 \left( \begin{array}{cccc} 
1 & -\tau_1 & -\tau_1\rho_1 & -\rho_1 \\
-\tau_1 & |\tau |^2 & \rho_1 |\tau |^2 & \tau_1 \rho_1 \\
-\tau_1 \rho_1 & \rho_1 | \tau |^2 & | \tau |^2 | \rho |^2 & 
\tau_1 | \rho |^2 \\
-\rho_1 & \tau_1 \rho_1 & \tau_1 | \rho |^2 & |\rho |^2 
\end{array} \right)  .
\end{equation}

An element $C$ of the Lie algebra of ${\rm O}(n,n)$ 
satisfies the constraint 
\begin{equation}
\label{lieconstraint}
\eta C+C^T \eta =0  .
\end{equation}
We are free to choose any form for $C$ when reducing 
the theory on a circle with a duality twist
as long as it belongs to this Lie algebra. 
We specify
\begin{equation}
\label{defCO22}
C = \left( \begin{array}{cc} 
m_1 J & m_3 J \\
m_3 J & m_1 J \end{array} \right)  ,
\end{equation} 
where $m_{1,3}$ are constant mass parameters and $J$ is the ${\rm SL}(2,R)$ 
metric defined in Eq. (\ref{Celliptic}) 
and satisfying $J^2 = -{\rm I}_2$. 
After a twisted reduction with this mass matrix, the moduli 
potential takes the form  
\begin{equation}
\label{modpotO22}
U= -2 (m_1^2+m_3^2)-
\frac{1}{2} {\rm Tr} \left( C \eta {\cal{M}} \eta C {\cal{M}} 
\right)
\end{equation}
and, since $C=-C^T$, this potential admits 
a global Minkowski minimum at ${\cal{M}}_0 = {\rm I}_4$. 

The ${\rm SL}(2,R)$ subgroups of ${\rm O}(2,2)$ can be 
made more transparent by defining the ${\rm SL}(2,R)$ matrices
\begin{equation}
\label{SG2}
S \equiv \frac{1}{\tau_2} \left( \begin{array}{cc}
1 & -\tau_1 \\
-\tau_1 & |\tau |^2 \end{array} \right)  
, \qquad 
P \equiv \frac{1}{\rho_2} \left( \begin{array}{cc}
1 & -\rho_1 \\
-\rho_1 & |\rho |^2 \end{array} \right)  .
\end{equation}
The moduli matrix (\ref{matrixM1}) 
can then be written in the block form
\begin{equation}
\label{blockform}
{\cal{M}} = \frac{1}{\rho_2} \left( \begin{array}{cc} 
S & -\rho_1 SJ \\
\rho_1 JS & |\rho |^2 S^{-1} \end{array} \right)
\end{equation}
and this implies that 
the moduli potential (\ref{modpotO22}) simplifies to 
\begin{equation}
\label{modpotSL2}
U= m_1^2 \, {\rm Tr} \left( - {\rm I}_2 + S^2 \right)
+ m_3^2 \, {\rm Tr} \left( - {\rm I}_2 + P^2 \right)  .
\end{equation}

The potential (\ref{modpotSL2}) is manifestly invariant under two global 
${\rm SO}(2)$ transformations. The ${\rm O}(2,2)$ matrix that 
generates such transformations is given by 
\cite{blsw1993}:
\begin{equation}
\label{defOmega}
\Omega \equiv \Omega_{\tau} \Omega_{\rho} =
\left( \begin{array}{cc}
c_1 \Sigma^T  & s_1 \Sigma^T J\\
s_1 \Sigma^T J & c_1 \Sigma^T 
\end{array} \right)  ,
\end{equation}
where 
\begin{eqnarray}
\label{omegarho}
\Omega_{\rho} \equiv \left( \begin{array}{cc}
c_1 {\rm I}_2 & s_1 J \\
s_1 J & c_1 {\rm I}_2 \end{array} \right) 
\, \qquad 
\Omega_{\tau} \equiv
\left( \begin{array}{cc} 
\Sigma^T & 0 \\
0 & \Sigma^T \end{array} \right) 
\nonumber \\
\Sigma \equiv 
\left( \begin{array}{cc}
c_2 & s_2 \\
-s_2 & c_2 \end{array} \right)
\end{eqnarray}
and $c_i \equiv \cos \theta_i$, $s_i \equiv \sin \theta_i$. 
Matrix (\ref{defOmega}) satisfies the conditions 
$\Omega \Omega^T = {\rm I}_4$  
and $\Omega^T \eta  \Omega =\eta$ and these conditions imply that 
$\eta \Omega = \Omega \eta$. Moreover, it may be verified that 
$C\Omega = \Omega C$, which implies that  
$C \eta = \Omega C \eta \Omega^T$. Hence, 
the moduli potential (\ref{modpotO22})
is invariant the global symmetry transformation 
\begin{equation}
\label{symtrans}
\tilde{\cal{M}} = \Omega^T {\cal{M}} \Omega
\end{equation}
and the dimensionally reduced action will respect this 
symmetry, where the metric transforms as a singlet. 

We may now employ the arguments of Section \ref{sec:noncan} to 
deduce that the potential (\ref{modpotO22}) 
admits cosmological scaling behaviour in four 
dimensions. Eq. (\ref{modpotSL2}) implies that the real (axionic) 
components of the complex moduli (\ref{taurho}) 
appear quadratically in the potential and these degrees of freedom 
may therefore be set to zero. In this case the potential  
reduces to 
\begin{equation}
\label{doublepot}
V= 4 e^{\sqrt{3} \varphi} \sum_{j=1,3} 
m^2_j  \, {\rm sinh}^2 \phi_j  ,
\end{equation}
where we have labelled the canonical fields 
$ e^{-\phi_1} = \sqrt{\Gamma}/\Gamma_{nn}$ 
and $e^{-\phi_2} = \sqrt{\Gamma}$, respectively.  
Eq. (\ref{doublepot}) reduces to Eq. (\ref{toypot}) 
in the limit $|\phi_j| \ll 1$ and the cosmological scaling solution 
(\ref{fluidscale}) will therefore 
arise in the presence of a background fluid, since  
this behaviour is independent of the mass parameters $m_j$ and the number of 
moduli fields $\phi_j$. Moreover, the symmetry transformation
(\ref{symtrans}) may be used to generate 
non-trivial, real components for the complex moduli fields (\ref{taurho}). 
Specifically, the matrix $\Omega_{\rho}$ generates the 
transformation
\begin{equation}
\label{rhotransformation}
\tilde{\rho} =\frac{s_1 + c_1\rho}{c_1 -s_1\rho}  , 
\qquad \tilde{\tau} =\tau
\end{equation}
that leaves the complex scalar field $\tau$ 
invariant and the matrix $\Omega_{\tau}$ 
generates the transformation
\begin{equation}
\label{tautransformation}
\tilde{\tau}=\frac{c_2 \tau -s_2}{s_2 \tau +c_2} , \qquad 
\tilde{\rho} =\rho
\end{equation}
that leaves $\rho$ invariant. Since the four-dimensional 
metric is invariant under these transformations, 
the scaling behaviour (\ref{fluidscale}) 
will be preserved when the axion fields evolve non-trivially. 

In the following Section, we extend our discussion to 
consider models which exhibit a global ${\rm SL}(n,R)$ symmetry. 

\section{Scaling Cosmologies from a ${\mathbf{G=SL(n,R)}}$ Duality Twist}

\label{sec:torus}

The ${\rm SL}(n,R)$ non-linear sigma-model 
is motivated by the Kaluza-Klein compactification 
of pure Einstein gravity on an $n$-dimensional torus\footnote{The case 
$n=4$ is of particular interest, since ${\rm SL}(4,R)$ is isomorphic 
to ${\rm SO}(3,3)$. This latter group arises as the global symmetry 
of the dimensionally reduced NS-NS string effective action on $T^3$.}. 
The dilatons which arise in such a reduction parametrize a geodesically 
complete submanifold in the coset space and this allows for a consistent 
truncation where the axion fields are frozen. 
We therefore consider the maximal Abelian subgroup of ${\rm SL}(n,R)$, 
where the moduli matrix ${\cal{M}}$ takes a diagonal form
\begin{equation} 
\label{diagM}
\hat{{\cal{M}}}={\rm diag} \left( e^{ -\vec{\beta}_a  \cdot \vec{\phi}}
\right) , \qquad a=1, \ldots , n
\end{equation}
and the vectors $\vec{\beta}_a$ denote the weights of 
the ${\rm SL}(n,R)$-algebra and satisfy the conditions 
\begin{eqnarray}
\label{defweights}
\sum_a \beta_{ai} =0 
\nonumber \\
\sum_a \beta_{ai}\beta_{aj} 
= 2\delta_{ij} 
\nonumber 
\\
\vec{\beta}_a \cdot \vec{\beta}_b  
= 2 \delta_{ab} - \frac{2}{n}
\end{eqnarray}
in the fundamental representation.  
It follows immediately from Eq. (\ref{defweights}) 
that the kinetic sector of the $(n-1)$ moduli fields   
takes the canonical form:  
\begin{equation}
{\rm Tr} \left( \nabla \hat{{\cal{M}}} \nabla 
\hat{{\cal{M}}}^{-1} \right) = -2 \delta_{ab} 
\nabla \hat{\phi}^a \nabla \hat{\phi}^b .
\end{equation}

The matrix $C$ should belong to the Lie algebra of ${\rm SL}(n,R)$, i.e, 
it should be a real, traceless, $n \times n$ matrix. We consider the 
case where 
\begin{equation}
\label{defC}
C = \left( \begin{array}{cccccc}
0 & m_1 & 0 & 0 & 0 & 0  \\
-m_1 & 0 & 0 & 0 & 0 & 0 \\
0 & 0 & 0 & m_3 & 0 & 0  \\
0 & 0 & -m_3 & 0 & 0 & 0 \\
0 & 0 & 0 & 0 & 0 & m_5  \\
0 & 0 & 0 & 0 & -m_5 & \ddots \\ 
\end{array} \right)
\end{equation}
and $m_a$ are constants. 
We reduce the $(D+1)$-dimensional 
${\rm SL}(n,R)/{\rm SO}(n)$ non-linear sigma-model on a circle with 
a duality twist by employing Eqs. (\ref{diagM}) and (\ref{defC}). 
When $n$ is even, we find that the action reduces to that of 
Eq. (\ref{twistedaction}), where the 
potential (\ref{twistedpot}) is given by 
\begin{equation}
\label{starpot}
V=2  e^{-2(D-1)\alpha \psi} 
\sum_j m_j^2   \sinh^2 \left[  \frac{1}{2} \left(  \vec{\beta}_{j+1}-
\vec{\beta}_j \right) \cdot \vec{\phi} \right]  
\end{equation}
and the sum is over $j=1, 3, \ldots, (n-1)$. 

Let us consider the case $n=6$ as a concrete example. 
A convenient basis for the weights of ${\rm SL}(6,R)$ which satisfies 
the constraints (\ref{defweights}) is given by 
\begin{eqnarray}
\vec{\beta}_1 =  \left( 1, \frac{1}{\sqrt{3}} , 
\frac{1}{\sqrt{6}}, \frac{1}{\sqrt{10}}, \frac{1}{\sqrt{15}} \right) 
, \qquad 
\vec{\beta}_2 =  \left( -1, \frac{1}{\sqrt{3}} , 
\frac{1}{\sqrt{6}}, \frac{1}{\sqrt{10}}, \frac{1}{\sqrt{15}} \right) 
\nonumber 
\\
\vec{\beta}_3 =  \left( 0, -\frac{2}{\sqrt{3}} , 
\frac{1}{\sqrt{6}}, \frac{1}{\sqrt{10}}, \frac{1}{\sqrt{15}} \right) 
, \qquad 
\vec{\beta}_4 =  \left( 0, 0, 
-\frac{3}{\sqrt{6}}, \frac{1}{\sqrt{10}}, \frac{1}{\sqrt{15}} \right) 
\nonumber 
\\ 
\vec{\beta}_5 =  \left( 0, 0, 
0, -\frac{4}{\sqrt{10}}, \frac{1}{\sqrt{15}} \right) 
, \qquad 
\vec{\beta}_6 =  \left( 0, 0, 
0, 0, -\frac{5}{\sqrt{15}} \right) 
\label{weightbasis}
\end{eqnarray}
and it follows from Eq. (\ref{weightbasis}) that
\begin{eqnarray}
\vec{\beta}_2 - \vec{\beta}_1 = (-2, 0,0,0,0) 
, \qquad 
\vec{\beta}_4 - \vec{\beta}_3 = \left( 0, \frac{2}{\sqrt{3}}, 
-\frac{4}{\sqrt{6}} , 0, 0 \right) \nonumber \\
\vec{\beta}_6 - \vec{\beta}_5 = \left( 0, 0, 0, \frac{4}{\sqrt{10}} , 
-\frac{6}{\sqrt{15}} \right)  ,
\end{eqnarray}
respectively. 
Consequently, the two ${\rm SO}(2)$ field redefinitions
\begin{eqnarray}
\psi_3 = \frac{1}{2} \left( \frac{2}{\sqrt{3}} \phi_2 
- \frac{4}{\sqrt{6}} \phi_3 \right)
, \qquad 
\gamma_3 = \frac{1}{2} \left( \frac{4}{\sqrt{6}} \phi_2 
+ \frac{2}{\sqrt{3}} \phi_3 \right)
\nonumber \\
\psi_5 = \frac{1}{2} \left( \frac{4}{\sqrt{10}} \phi_4 - \frac{6}{\sqrt{15}} 
\phi_5 \right) , \qquad 
\gamma_5 = \frac{1}{2} \left( \frac{6}{\sqrt{15}} \phi_4 
+ \frac{4}{\sqrt{10}}
\phi_5 \right)
\end{eqnarray}
decouple the scalar degrees of freedom $\gamma_3$ and $\gamma_5$ 
from the potential (\ref{starpot}). These fields may then be set to zero.  
A further toroidal compactification 
to four dimensions, as outlined in Section \ref{sec:twist}, 
then results in an action consisting of 
four canonical fields which interact through the potential   
\begin{equation}
\label{finalpot}
V = 2 e^{\sqrt{3} \varphi} \sum_{j=1,3,5} m_j^2 \, 
{\rm sinh}^2 \, \psi_j  ,
\end{equation}
where we have labelled $\phi_1 = - \psi_1$. 

The potential (\ref{finalpot}) is of the general form (\ref{genpot}).
It belongs to the type II class of models since the seven 
$\vec{\alpha}_a$ vectors are linearly dependent and 
satisfy the affine relation (\ref{affine}). 
Moreover, the rank of the $7 \times 4$ matrix formed from the  
$\vec{\alpha}_a$ is four, which implies that no further decoupling 
of the fields from the potential can be made. Since the potential 
(\ref{finalpot}) is in the separable form (\ref{seppot}) 
and reduces to Eq. (\ref{toypot}) in the small $\psi_j$ limit, 
we may conclude immediately that this type of twisted compactification 
admits the cosmological scaling solution (\ref{fluidscale}). 

It can be verified that since the 
weights of ${\rm SL}(n,R)$ satisfy the constraints 
(\ref{defweights}), similar potentials 
of the form (\ref{finalpot}) arise for all even
values of $n$ and differ only by the number of hyperbolic terms 
in the summation. Scaling behaviour will therefore  
be generated in these cases, since such 
behaviour is independent of the number of oscillating fields
and the magnitude of the mass parameters $m_j$.  

Likewise, the scaling cosmologies are preserved in the 
presence of the axionic degrees of freedom. In general, 
when the mass matrix belonging to the Lie algebra of $G$ is antisymmetric, 
the monodromy ${\cal{C}} = 
\exp (C)$ is an element of the rotation group ${\rm O}(n)$ and therefore 
satisfies ${\cal{C}}{\cal{C}}^T = {\rm I}_n$. 
This implies that a Minkowski minimum in the potential 
located at ${\cal{M}}_0 = {\rm I}_n$ in the moduli space 
is a fixed point under the action of the 
monodromy group, since 
${\cal{M}}_0 \rightarrow {\cal{C}}^T {\cal{M}}_0 {\cal{C}} 
= {\cal{C}}^T{\cal{C}} = {\rm I}_n$ \cite{dh2002}. Consequently, 
the reduced theory 
will be symmetric under the action of an ${\rm SO}(n)$ transformation
if $G' = {\rm SO}(n)$ is a subgroup of $G$.  
In the case where $G={\rm SL}(n,R)$, such 
a transformation may be employed to generate a cosmological 
background with non-trivial axion fields, as was demonstrated 
explicitly  in Section \ref{sec:noncan} when $n=2$. More generally, the moduli 
potential (\ref{twistedpot}) is invariant under ${\cal{M}} \rightarrow 
\Omega^T {\cal{M}} \Omega$ if 
the mass matrix transforms as $C\rightarrow \Omega^T C \Omega$, where  
$\Omega \Omega^T={\rm I}_n$. In this case,  
$\Omega^T C\Omega $ belongs to the Lie algebra of 
${\rm SL} (n,R)$ if ${\rm Tr} \, C =0$. 

In the following Section, we use the above results to 
resolve an apparent ambiguity that arose in the work of Ref. \cite{ksh2007}.

\section{Compactification on an Elliptic Twisted Torus}

\label{sec:ksh}

When the duality symmetry has a geometrical origin, as is the case when 
$G={\rm SL}(n,R)$, a reduction with 
a duality twist is equivalent to compactification on a twisted torus 
\cite{hre2005}. In this latter scheme,  
the dependence of the fields on the internal coordinates $y^i$ 
arises through a matrix $\sigma^i_m (y)$ with inverse $\sigma^m_i (y)$
\cite{ss1979}. The compactification is consistent if the matrices 
$\sigma^i_m(y)$ satisfy the constraint  
${f^m}_{np} =-\sigma^i_n \sigma^j_p ( \partial_i \sigma^m_j - \partial_j
\sigma^m_i)$, where  ${f^m}_{np}$ are the structure constants for a Lie 
group $G$. The four-dimensional potential is then determined 
by these structure constants and the metric on the internal manifold.

The elements of the mass matrix (\ref{defC}) represent the 
structure constants of the elliptic twisted torus employed by KSH 
in the compactification of eleven-dimensional pure Einstein gravity 
\cite{ksh2007}. This manifold is related to the Lie group ${\rm ISO}(n)$, 
corresponding to the group of isometries of $n$-dimensional Euclidean space.  
In the case where the internal manifold has a diagonal metric, 
the potential in the twisted torus compactification was found to have the 
form \cite{ksh2007}  
\begin{eqnarray}
\label{elliptictorus}
V(\varphi_a) = e^{-\frac{3}{\sqrt{7}} \varphi_1} \left[ 
\right. 
\nonumber \\
\left. 
m_1^2 
e^{-\frac{2}{\sqrt{3}} \varphi_3 + \frac{\varphi_4}{\sqrt{6}} + 
\frac{\varphi_5}{\sqrt{10}} + \frac{\varphi_6}{\sqrt{15}} + 
\frac{\varphi_7}{\sqrt{21}}} \left( e^{\varphi_2} - e^{\sqrt{3} \varphi_3}
\right)^2
\right. \nonumber \\
\left. 
+ m_3^2  e^{\varphi_2 + \frac{\varphi_3}{\sqrt{3}} - \sqrt{\frac{2}{3}} 
\varphi_4 - \sqrt{\frac{8}{5}}\varphi_5 + \frac{\varphi_6}{\sqrt{15}}
+ \frac{\varphi_7}{\sqrt{21}}} \left( 
e^{\sqrt{\frac{3}{2}} \varphi_4} - e^{\sqrt{\frac{5}{2}} \varphi_5} 
\right)^2
\right. \nonumber \\
\left. 
+ m_5^2 e^{\varphi_2 + \frac{\varphi_3}{\sqrt{3}} + 
\frac{\varphi_4}{\sqrt{6}} + \frac{\varphi_5}{\sqrt{10}}
- 4 \frac{\varphi_6}{\sqrt{15}}
-2 \sqrt{\frac{3}{7}} \varphi_7} \left( 
e^{\sqrt{\frac{5}{3}} \varphi_6} - e^{\sqrt{\frac{7}{3}} \varphi_7} 
\right)^2 \right]  ,
\end{eqnarray}
where the seven moduli fields $\varphi_a$ are related to the 
components of the internal metric  and are canonically coupled to 
Einstein gravity. The field $\varphi_1$ 
parametrizes the volume of the internal space. 

The numerical results of \cite{ksh2007} indicate that at late-times 
the system oscillates around the vacuum $\varphi_2 = \sqrt{3} \varphi_3$, 
$\varphi_4 = \sqrt{5/3} \varphi_5$ and $\varphi_6 = \sqrt{7/5} \varphi_7$, 
where the modulus field $\varphi_1$ increases monotonically with time. This 
would imply that the potential should reduce to Eq. (\ref{toypot}) 
in the neighbourhood of this critical point, where $c= - 3/\sqrt{7}$. 
Consequently, the scalar fields should contribute a fraction $\Omega_s 
\simeq 7(\gamma -1)/3$ of the total energy density during 
the scaling regime. However, although the numerical simulations of 
\cite{ksh2007} indeed confirmed the existence 
of scaling behaviour, it was found that $\Omega_s \simeq 1/3$ 
in the presence of a radiation fluid, which 
corresponds to $c=\sqrt{3}$. The results from Section \ref{sec:torus}
imply that this is indeed the correct value for the coupling parameter. 

To gain further insight, we rewrite the potential (\ref{elliptictorus}) 
in the more compact form 
\begin{equation}
V = \sum_{j=1}^6 m^2_j e^{\vec{\alpha}_j \cdot \vec{\varphi}} - 
2 M^2 e^{\vec{\alpha}_7 \cdot 
\vec{\varphi}}  ,
\end{equation}
where $\vec{\varphi} = (\varphi_1 , \ldots , \varphi_7 )$, the constants 
$m_1=m_2$, $m_3=m_4$, $m_5=m_6$ and $M^2\equiv m_1^2+m_3^2+m_5^2$. 
The constant vectors $\vec{\alpha}_i$ parametrizing the field 
couplings are given by  
\begin{eqnarray}
\vec{\alpha}_1 = \left( -\frac{3}{\sqrt{7}}, 2, -\frac{2}{\sqrt{3}}, 
\frac{1}{\sqrt{6}} , \frac{1}{\sqrt{10}}, \frac{1}{\sqrt{15}} , 
\frac{1}{\sqrt{21}} \right) 
\nonumber
\\
\vec{\alpha}_2 = \left( -\frac{3}{\sqrt{7}}, 0, \frac{4}{\sqrt{3}}, 
\frac{1}{\sqrt{6}} , \frac{1}{\sqrt{10}}, \frac{1}{\sqrt{15}} , 
\frac{1}{\sqrt{21}} \right) 
\nonumber
\\
\vec{\alpha}_3 = \left( -\frac{3}{\sqrt{7}}, 1, \frac{1}{\sqrt{3}}, 
\sqrt{\frac{8}{3}} , -\sqrt{\frac{8}{5}}, \frac{1}{\sqrt{15}} , 
\frac{1}{\sqrt{21}} \right) 
\nonumber
\\
\vec{\alpha}_4 = \left( -\frac{3}{\sqrt{7}}, 1, \frac{1}{\sqrt{3}}, 
-\sqrt{\frac{2}{3}} , \sqrt{\frac{18}{5}}, \frac{1}{\sqrt{15}} , 
\frac{1}{\sqrt{21}} \right) 
\nonumber
\\
\vec{\alpha}_5 = \left( -\frac{3}{\sqrt{7}}, 1, \frac{1}{\sqrt{3}}, 
\frac{1}{\sqrt{6}} , \frac{1}{\sqrt{10}}, \frac{6}{\sqrt{15}} , 
-\frac{6}{\sqrt{21}} \right) 
\nonumber
\\
\vec{\alpha}_6 = \left( -\frac{3}{\sqrt{7}}, 1, \frac{1}{\sqrt{3}}, 
\frac{1}{\sqrt{6}} , \frac{1}{\sqrt{10}}, -\frac{4}{\sqrt{15}} , 
\frac{8}{\sqrt{21}} \right) 
\nonumber
\\
\vec{\alpha}_7 = \left( -\frac{3}{\sqrt{7}}, 1, \frac{1}{\sqrt{3}}, 
\frac{1}{\sqrt{6}} , \frac{1}{\sqrt{10}}, \frac{1}{\sqrt{15}} , 
\frac{1}{\sqrt{21}} \right) 
\label{thealphas}
\end{eqnarray}
and satisfy the affine relations: 
\begin{equation}
\label{convexconditions}
\vec{\alpha}_j+\vec{\alpha}_{j+1} = 2\vec{\alpha}_7  , \qquad j =1,3,5,
\end{equation}
Hence, the potential (\ref{elliptictorus}) is of the type II 
discussed in Section \ref{sec:approx}. 

The $7 \times 7$ matrix formed from the 
$\vec{\alpha}_j$ vectors has rank $R=4$, so three of the scalar 
fields can be decoupled from the potential. 
We now define new fields such that  
$\vec{\varphi} \equiv \varphi \vec{n} + \vec{\varphi}_{\perp}$,
where $\vec{n}$ is a unit vector, $\varphi$ is a scalar function 
and $\vec{\varphi}_{\perp}$ is perpendicular 
to $\vec{n}$, i.e., $\vec{n}\cdot \vec{\varphi}_{\perp} =0$
\cite{gl1999,lp1995}. The projections of the 
vectors $\vec{\alpha}_j$ onto $\vec{n}$ must satisfy the constraint 
$\vec{\alpha}_j \cdot \vec{n} = c$ for all $j$, where $c$ is a constant. 
The value of $c$ may be deduced by considering the symmetric matrix
\begin{equation}
\label{matrix}
A_{ij} \equiv \vec{\alpha}_i \cdot \vec{\alpha}_j = 
\left( \begin{array}{ccccccc}
7 & -1 & 3 & 3 & 3 & 3 & 3 \\
-1 & 7 & 3 & 3 & 3 & 3 & 3 \\
3 & 3 & 7 & -1 & 3 & 3 & 3 \\
3 & 3 & -1 & 7 & 3 & 3 & 3 \\
3 & 3 & 3 & 3 & 7 & -1 & 3 \\
3 & 3 & 3 & 3 & -1 & 7 & 3 \\
3 & 3 & 3 & 3 & 3 & 3 & 3 
\end{array} 
\right) .
\end{equation}
Since $\vec{\alpha}_j \cdot \vec{\alpha}_7 =3$ 
for all $j=1, \ldots , 7$, we may specify  
$\vec{n}={\vec{\alpha}_7}/{\sqrt{3}}$ and $c= \sqrt{3}$. 

The equations of motion for the $\vec{\varphi}$-fields are given by  
\begin{equation}
\label{scalareom}
\partial^2 \vec{\varphi}- \sum_{j=1}^6 m_j^2 
\vec{\alpha}_j e^{\vec{\alpha}_j \cdot 
\vec{\varphi}} +2M^2\vec{\alpha}_7 e^{\vec{\alpha}_7 \cdot  \vec{\varphi}}
=0 .
\end{equation}
Taking the dot product of Eq. (\ref{scalareom}) with 
respect to the  unit vector $\vec{n}$ yields
\begin{equation}
\label{varphieom}
\partial^2 \varphi -ce^{c\varphi} \left[ 
\sum_{j=1}^6 m_j^2 e^{\vec{\alpha}_j \cdot \vec{\varphi}_{\perp}}
-2M^2e^{\vec{\alpha}_7 \cdot \vec{\varphi}_{\perp}} \right] =0
\end{equation}
and it follows after substitution of 
Eq. (\ref{varphieom}) into the field equations 
(\ref{scalareom}) that
\begin{equation}
\label{simplifyperpeom}
\partial^2 \vec{\varphi}_{\perp} -e^{c \varphi} 
\sum_{j=1}^6 m_j^2 \left( \vec{\alpha}_j 
- \vec{\alpha}_7 \right) e^{\vec{\alpha}_j \cdot 
\vec{\varphi}_{\perp}}  =0  .
\end{equation}
Employing Eq. (\ref{thealphas}) for each of the 
components $\varphi_{\perp i}$ of $\vec{\varphi}_{\perp}$ 
in Eq. (\ref{simplifyperpeom}) then implies that 
\begin{eqnarray}
\label{massless}
\partial^2 \left( \varphi_{\perp 2} + \frac{\varphi_{\perp 3}}{\sqrt{3}} 
\right) =0 
, \qquad 
\partial^2 \left(\frac{\varphi_{\perp 4}}{\sqrt{3}} + 
\frac{\varphi_{\perp 5}}{\sqrt{5}} \right) =0
\nonumber \\
\partial^2 \left( \frac{\varphi_{\perp 6}}{\sqrt{5}} + 
\frac{\varphi_{\perp 7}}{\sqrt{7}} \right) =0  .
\end{eqnarray}
These linear combinations of $\varphi_{\perp i}$
represent the three degrees of freedom 
that decouple from the potential. 

The field equations for three of the remaining four 
interacting fields can be derived by 
considering the dot product $\vec{\alpha}_i \cdot \vec{\varphi}_{\perp}$
with $i=1, \ldots , 6$. Only three of the 
$\vec{\alpha}_i \cdot \vec{\varphi}_{\perp}$ fields are independent, since the 
orthogonality constraint $\vec{\alpha}_7 \cdot \vec{\varphi}_{\perp} =0$ 
and the affine conditions (\ref{convexconditions}) together imply that 
\begin{equation}
\label{cjconstraints}
\vec{\alpha}_i \cdot \vec{\varphi}_{\perp} 
= -  \vec{\alpha}_{i+1} \cdot \vec{\varphi}_{\perp} , \qquad 
i=1,3,5.
\end{equation}
Hence, taking the dot product of Eq. (\ref{simplifyperpeom})
with $\vec{\alpha}_j$  results in the equations of motion: 
\begin{equation}
\label{cperpeom}
\partial^2 \left( \vec{\alpha_i} \cdot \vec{\varphi}_{\perp} \right)
-e^{c\varphi}  \sum_{j=1}^6 m_j^2 \left( 
\vec{\alpha}_i \cdot \vec{\alpha}_j -3 \right) e^{\vec{\alpha}_j \cdot 
\vec{\varphi}_{\perp}}  =0
\end{equation}
and it follows from the matrix (\ref{matrix}) and constraints 
(\ref{cjconstraints}) that
\begin{equation}
\label{producteom}
\partial^2 \left( \vec{\alpha_i} \cdot \vec{\varphi}_{\perp} \right)
-8 m_i^2 e^{c\varphi} \, {\rm sinh} \, 
\left( \vec{\alpha}_i \cdot \vec{\varphi}_{\perp}
\right) =0 . 
\end{equation}
Finally, the fourth interacting scalar is the function 
$\varphi$ and its field equation (\ref{varphieom}) reduces to 
\begin{equation}
\label{simplevarphieom}
\partial^2 \varphi -4ce^{c \varphi} \sum_{j=1,3,5} m_j^2 
\, {\rm sinh}^2 \, \left( 
\frac{\vec{\alpha}_j \cdot \vec{\varphi}_{\perp}}{2} \right)
=0  .
\end{equation}

The question that now arises is whether the $\vec{\alpha}_j \cdot 
\vec{\varphi}_{\perp}$ fields are canonical. In general, the field rotation 
introduces off-diagonal terms into the metric 
of the field space. On the other hand, a consistent solution to 
Eq. (\ref{massless}) is given by 
\begin{equation}
\label{consistentsol}
\varphi_{\perp 2} = - \frac{1}{\sqrt{3}} \varphi_{\perp 3} , 
\qquad \varphi_{\perp 4} = - \sqrt{\frac{3}{5}} \varphi_{\perp 5}
, \qquad \varphi_{\perp 6} =-\sqrt{\frac{5}{7}} \varphi_{\perp 7}
\end{equation}
and, since Eq. (\ref{massless}) and the constraint 
$\vec{\alpha}_7 \cdot \vec{\varphi}_{\perp} =0$ imply that 
$\varphi_{\perp 1} =0$, Eq. (\ref{consistentsol}) 
is equivalent to   
\begin{equation}
\label{cjconsistent}
\vec{\alpha}_1 \cdot \vec{\varphi}_{\perp} = 4 \varphi_{\perp 2} , 
\qquad 
\vec{\alpha}_3 \cdot \vec{\varphi}_{\perp} = \frac{8}{\sqrt{6}} 
\varphi_{\perp 4} , 
\qquad 
\vec{\alpha}_5 \cdot \vec{\varphi}_{\perp} = \frac{12}{\sqrt{15}} 
\varphi_{\perp 6}  .
\end{equation}
This implies that 
\begin{equation}
\label{kineticterms}
\partial \vec{\varphi} \cdot \partial \vec{\varphi}
= \left( \partial \varphi \right)^2 
+ \frac{1}{4} \sum_{j=1,3,5}  \left[ \partial \left( 
\vec{\alpha}_j \cdot  \vec{\varphi}_{\perp} \right) \right]^2 
\end{equation}
and we conclude, therefore, that the field equations
(\ref{producteom}) and (\ref{simplevarphieom}) can be
derived from the effective Lagrangian 
\begin{equation}
\label{Lagrangian}
{\cal{L}} = -\frac{1}{2} \partial \vec{\psi} \cdot \partial \vec{\psi}
-4  e^{c \varphi} \sum_{j=1,3,5} m_j^2 \, {\rm sinh}^2 \, \psi_j  ,
\end{equation}
where $\vec{\psi} \equiv (\varphi, \psi_1, \psi_3 ,\psi_5)$,
$\psi_j \equiv \frac{1}{2} \vec{\alpha}_j \cdot \vec{\varphi}_{\perp}$
and $c = \sqrt{3}$. The potential in Eq. (\ref{Lagrangian})  
is precisely of the form given in Eq. (\ref{finalpot}), 
confirming that the radiation scaling solution is indeed characterized by 
$\Omega_s = 1/3$.  

\section{Conclusion}

\label{sec:discuss}

In this paper we have investigated the cosmological 
consequences of the Scherk-Schwarz compactification 
of higher-dimensional non-linear sigma-models. If the 
$(D+1)$-dimensional theory exhibits a global 
symmetry $G$, reduction on a circle with a $G$ duality twist, followed 
by a Kaluza-Klein reduction on an isotropic $(D-4)$-torus,
results in an effective four-dimensional potential of the form 
$V = e^{c x} U ({\cal{M}})$, where $c = \sqrt{3}$ and 
${\cal{M}} \in G$ represents the matrix of moduli fields. 
The value of the coupling parameter $c = \sqrt{3}$ is independent of
the symmetry group $G$ and the spacetime dimensionality 
$D$. When the scalar fields are canonically coupled and the 
function $U$ admits a Minkowski ground state, oscillations of the 
moduli around this vacuum support a cosmological scaling 
solution\footnote{The solution (\ref{approxsol}) is only an 
approximate solution to the field equations and does not represent  
a global attractor of the system. Nonetheless, our numerical
results indicate that it accurately describes the cosmic 
dynamics  over a large number of e-foldings.} in the 
presence of a perfect fluid \cite{ksh2007}. When the fluid has a relativistic 
equation of state, the fluid and scalar field 
energy densities are predicted to be  
$\Omega_{\gamma}  =2 \Omega_s =2/3$ and this was confirmed by 
numerical simulations. The scaling solution (\ref{approxsol}) therefore 
provides a very good approximation to the cosmological dynamics in this 
regime.

These conclusions can be extended to non-canonical 
moduli when the dimensionally reduced action 
is symmetric under a global symmetry $G' \subset G$, where the metric 
transforms as a singlet under the action of $G'$. In the 
case where $G={\rm SL}(n,R)$, the moduli potential  
is invariant under $G'={\rm SO}(n)$. Since 
the perfect fluid source does not break the 
corresponding symmetry of the field equations, such a 
symmetry transformation may be employed to generate 
a cosmological scaling solution with dynamical, non-canonical fields 
from a background where such fields are trivial. 

Finally, we have resolved an apparent ambiguity 
between the analytical and numerical results of \cite{ksh2007}
by demonstrating explicitly that the potential derived 
from the reduction of eleven-dimensional pure Einstein gravity 
on an elliptic twisted torus reduces to a 
potential of the form (\ref{toypot}) with $c= \sqrt{3}$
after three of the moduli fields have been consistently decoupled
from the potential. 
 
\appendix

\section{Cosmological Field Equations for the 
${\mathbf{SL(2,R)}}$ Model}

The scalar field equations derived by extremizing the action
(\ref{fullaction}) are given by 
\begin{eqnarray}
\label{appendix1}
&&\ddot{\varphi}+3H\dot{\varphi}+\frac{\sqrt{3} m^2}{2} e^{\sqrt{3}{\varphi}}
\left[e^{2\phi}\left(\chi^2+1\right)^2+e^{-2\phi}
+2\left(\chi^2-1\right)\right]=0\,,\\
&&\ddot\phi+3H\dot\phi-e^{2\phi}\dot\chi^2+m^2 e^{\sqrt{3}{\varphi}}
\left[e^{2\phi}\left(\chi^2+1\right)^2-e^{-2\phi}\right]=0\,,\\
&&\ddot\chi+3H\dot\chi+2\dot\chi\dot\phi+2\chi m^2 e^{\sqrt{3}{\varphi}}
\left[\left(\chi^2+1\right)+e^{-2\phi}\right]=0\,.
\end{eqnarray}
The perfect fluid satisfies the standard conservation equation
\begin{equation}
\dot\rho_\gamma+3H\left(\rho_\gamma+p_\gamma\right)=0\,,\\
\end{equation}
and the Friedmann constraint equation takes the form 
\begin{eqnarray}
\label{appendix5}
H^2=\frac{1}{6}\left[\rho_\gamma
+\frac{1}{2}\dot{\varphi}^2+\frac{1}{2}\dot\phi^2
+\frac{1}{2}e^{2\phi}\dot\chi^2
\right. \nonumber \\
\left. 
+\frac{m^2}{2} e^{\sqrt{3}{\varphi}}
\left( e^{2\phi}\left(\chi^2+1\right)^2+e^{-2\phi}
+2\left(\chi^2-1\right)\right)
\right] \,.
\end{eqnarray}

\ack
We thank C. Hidalgo for helpful discussions. 

\section*{References}

\bibliographystyle{iopart-num-ih}

\bibliography{qmw07-4}

\end{document}